\documentclass[12pt,preprint]{aastex}

\shorttitle{Ellipticity of Spiral Galaxies}
\shortauthors{Ryden}

\begin{document}

\title{The Ellipticity of the Disks of Spiral Galaxies} 
\author{Barbara S. Ryden}
\affil{Department of Astronomy, The Ohio State University}
\affil{140 W. 18th Avenue, Columbus, OH 43210\\
     ryden@astronomy.ohio-state.edu
}

\begin{abstract}

The disks of spiral galaxies are generally elliptical rather
than circular. The distribution of ellipticities can be fit
with a log-normal distribution. For a sample of 12,764 galaxies
from the Sloan Digital Sky Survey Data Release 1 (SDSS DR1),
the distribution of apparent axis ratios in the $i$ band
is best fit by a log-normal distribution
of intrinsic ellipticities with
$\ln\varepsilon = -1.85 \pm 0.89$. For a sample of
nearly face-on spiral galaxies, analyzed by Andersen \&
Bershady using both photometric and spectroscopic data, 
the best fitting distribution of ellipticities has
$\ln\varepsilon = -2.29 \pm 1.04$. Given the small size
of the Andersen \& Bershady sample, the two distributions
are not necessarily inconsistent with each other.
If the ellipticity of the potential were equal to that of
the light distribution of the SDSS DR1 galaxies, it would
produce 1.0 magnitudes of scatter in the Tully-Fisher
relation, greater than is observed. The Andersen \& Bershady
results, however, are consistent with a scatter as small
as 0.25 magnitudes in the Tully-Fisher relation.

\end{abstract}


\section{Introduction}
\label{intro}

The disks of spiral galaxies are not axisymmetric structures.
Spiral arms are an obvious deviation from axisymmetry,
as are bars within barred spiral galaxies. In
addition, the overall shape of the disk may be elliptical
rather than circular. The shape of a stellar disk can be
roughly approximated as a triaxial spheroid
with principal axes of length $a \ge b \ge c$. A
typical stellar disk is relatively thin, with
$\gamma \equiv c/a \ll 1$, and mildly elliptical, with
$\varepsilon \equiv 1 - b/a \ll 1$. The exact distribution
of ellipticities for the disks of spiral galaxies has long
been a subject of debate.

Statistical statements about the distribution of
$\varepsilon$ can be made by looking at the distribution
of apparent axis ratio $q$ for a large population
of spiral galaxies. \citet{sa70} investigated a sample
of 254 spiral galaxies from the {\em Reference
Catalogue of Bright Galaxies} \citep{de64}. They
concluded that the observed axis ratios of the spiral
galaxies were consistent with their being a population of
circular disks, with typical thickness $\gamma \approx 0.25$.
\citet{bi81}, using data from the {\em Second Reference
Catalogue of Bright Galaxies} \citep{de76}, concluded
that late-type spirals were better fitted by mildly
elliptical disks ($\varepsilon = 0.1$) than by perfectly
circular disks. Using a sample of 13{,}482 spiral galaxies,
\citet{la92} found that the ellipticity was reasonably well
fitted by a Gaussian peaking at $\varepsilon = 0$:
\begin{equation}
f (\varepsilon) \propto \exp \left( - {\varepsilon^2
\over 2 \sigma_\varepsilon^2} \right) \ ,
\label{eq:apm}
\end{equation}
subject to the constraint $0 \leq \varepsilon \leq 1$. The best
fit was given by $\sigma_\varepsilon = 0.13$, yielding
an average ellipticity of $\langle \varepsilon \rangle
\approx 0.1$. The nonaxisymmetry of disks, which is signaled by a
scarcity of apparently circular galaxies, has been confirmed
by other photometric studies \citep{gr85, fa93, al02}.

Analyses which rely on the measured axis ratios of a large
sample of galaxies have certain intrinsic shortcomings.
First, they provide only a statistical statement about the
distribution of disk ellipticities in the sample examined;
they don't determine the axis ratio of any individual galaxy.
Moreover, since these studies rely purely on photometry, they
provide information about the ellipticity of the starlight
emitted by the galaxy. The ellipticity measured in this way
is affected by bars, eccentric rings and pseudorings,
loosely wound spiral arms, and patchy
star formation, and does not necessarily reflect the
overall ellipticity of the potential within which the
stars are orbiting. These intrinsic shortcomings are
avoided by methods using both photometric and kinematic
information.

Consider a disk of test particles orbiting in
a potential which is elliptical in the disk plane. The
closed particle orbits, when the potential ellipticity
is small, will be elliptical themselves \citep{bi78}. If the
potential is logarithmic, producing a flat rotation
curve, the ellipticity of the orbits will equal the
ellipticity of the potential. If the resulting elliptical
disk is seen in projection, the isophotal principal
axes will be misaligned with the kinematic principal
axes. (The kinematic principal axes can be defined as the
axis along which the line-of-sight component of the
rotation velocity is a maximum and the axis along which
is is zero.) Because of the misalignment, there is generally
a non-zero velocity gradient along the isophotal minor
axis, proportional to $\varepsilon \sin 2 \phi$, where
$\varepsilon$ is the ellipticity of the potential, and
$\phi$ is the azimuthal viewing angle measured relative
to the long axis of the potential \citep{fr92}. Studying
the velocity field of gas at large radii, where the dark
matter should dominate the potential, typically yields
$\varepsilon \sin 2 \phi \lesssim 0.07$ \citep{sc97}.

By combining integral-field spectroscopy with imaging data,
\citet{an01} were able to determine the intrinsic ellipticity
for a number of disk galaxies at low inclination. Since the
misalignment between the isophotal principal axes and the
kinematic principal axes decreases as the inclination increases,
the technique of \citet{an01} can only be usefully applied to
galaxies with inclination $i < 30^\circ$. For a sample of
28 spiral galaxies, \citet{an03} found that the intrinsic
ellipticities were well fitted by a log-normal distribution.
The mean ellipticity of the galaxies in their sample was
$\overline\varepsilon = 0.076$. The method of Andersen and
collaborators has its own shortcomings. The sample of galaxies
is relatively small. Systematic uncertainties in determining the
isophotal and kinematic major axes can mask the signal
produced by an elliptical disk \citep{ba03}. In addition, to
ensure that their sample contained only galaxies with small
inclination, \citet{an01} and \citet{an03} included only galaxies
that appeared nearly circular, with $q \geq 0.866$.
This selection bias discriminates against
intrinsically highly elliptical disks, which have a
relatively small probability of appearing nearly circular
in projection. Thus, the true distribution of ellipticities
will have a higher mean ellipticity and a larger
standard deviation than the Andersen-Bershady sample.

In this paper, I combine information from the two types of
analysis; the photometry-only method, which has the advantage
of large sample size, and the \citet{an01} method, which
has the advantage of including kinematic information which
probes the potential more directly. In section~\ref{sdss},
I use the measured apparent axis ratios of $\sim 12{,}800$ galaxies
from the Sloan Digital Sky Survey Data Release 1 to make
an estimate of their intrinsic disk ellipticities. In
section~\ref{andersen}, I reanalyze the disk ellipticities
found by \citet{an03}, taking into account the selection
bias, to find the underlying distribution of disk ellipticities.
I show that the two methods give results which are not
mutually inconsistent. In section~\ref{discuss}, I discuss
the implications of the disk ellipticity; in particular,
its relation to the scatter in the Tully-Fisher relation.

\section{The SDSS Sample}
\label{sdss}

The Sloan Digital Sky Survey (SDSS) is a digital photometric
and spectroscopic survey which will cover, when completed,
roughly one-quarter of the celestial sphere. Data Release 1
(DR1), issued to the astronomical community
in 2003 April, consists of 2099 square degrees of imaging
data in five photometric bands: $u$, $g$, $r$, $i$,
and $z$ \citep{fu96,st02}. The data
processing pipelines for DR1 (described in \citet{st02}
and \citet{ab03}) fit two models to the two-dimensional image
of each galaxy. The first model has a de Vaucouleurs profile
\citep{de48},
\begin{equation}
I(r) = I_0 \exp [ - 7.67 ( r / r_e )^{1/4} ] \ ,
\label{eq:deV}
\end{equation}
which is truncated beyond $7 r_e$, going smoothly to zero
at $8 r_e$; the model has some softening within $r_e/50$.
The second model has an exponential profile,
\begin{equation}
I(r) = I_0 \exp [ - 1.68 (r/r_e) ] \ ,
\label{eq:exp}
\end{equation}
which is truncated beyond $3 r_e$, going smoothly to zero
at $4 r_e$. Both models are assumed to have concentric
elliptical isophotes with constant position angle $\varphi_m$
and axis ratio $q_m$. Before each model is fitted to the
data, the model is convolved with a double Gaussian fitted
to the point-spread function (psf). Assessing each model with
a $\chi^2$ fit gives $r_e$, $q_m$, and $\varphi_m$ for the
best-fitting model, as well as $P ({\rm deV})$ and $P ({\rm exp})$,
the likelihood associated with the best de Vaucouleurs
model and the best exponential model, respectively. In
addition, each candidate galaxy is fitted directly with the
point-spread function, yielding $P ({\rm psf})$, the likelihood
associated with the psf fit.

To build a sample of disk galaxies from the SDSS DR1,
I selected objects flagged as galaxies whose model fits
in the $r$ band satisfied the criteria
$P ({\rm exp}) > 2 P ({\rm deV})$ and
$P ({\rm exp}) > 2 P ({\rm psf})$. In addition, each
galaxy was required to have a spectroscopic
redshift $z < 0.1$, to reduce the possibility of weak
lensing distorting the observed shape, and $z > 0.002$,
to eliminate a few contaminating foreground objects.
The spectroscopic sample of the SDSS DR1 contained 30,184
galaxies satisfying these criteria. Note that I have
selected galaxies on the basis of their surface brightness
profile (exponential rather than de Vaucouleurs), and
not on other criteria such as color or isophote shape.
It is generally true, though, that bright
early-type galaxies (elliptical and S0) are
better fitted by de Vaucouleurs profiles than by exponential
profiles, and that spiral galaxies are better fitted by exponential
profiles \citep{st01}.
Moreover, the spectroscopic sample of the DR1 contains
galaxies which are too high in surface brightness to be dwarf
ellipticals, which can also have exponential surface brightness
profiles \citep{bi91}. Thus, I will assume that the galaxies
which I have chosen on the basis of their exponential profiles
are disk-dominated spiral galaxies.

The model fits to the exponential galaxies provide one measure
of the axis ratio $q$. However, a model with constant position angle
$\varphi$ and axis ratio $q$ is not a good approximation to a real
spiral galaxy. Fortunately, the SDSS DR1 photometric analysis also
provides more robust, model-free measures of the axis ratio.
One useful shape measure is $q_{\rm am}$, the axis ratio determined
by the use of adaptive moments. In general, the technique of
adaptive moments determines the $n$th order moments of an image
using an elliptical weight function whose shape matches that
of the object \citep{be02, hi03}. In practice, the SDSS adaptive
moments use a Gaussian weight function $w(x,y)$ which is
matched to the size and ellipticity of the galaxy image $I(x,y)$.
With the weight function $w(x,y)$ known, the moments can be
calculated:
\begin{equation}
\vec{x_0} = {\int {\vec x} w(x,y) I(x,y) dx dy \over
\int w(x,y) I(x,y) dx dy} \ ,
\end{equation}
\begin{equation}
M_{xx} = {\int (x-x_0)^2 w(x,y) I(x,y) dx dy \over
\int w(x,y) I(x,y) dx dy} \ ,
\end{equation}
and so forth. The SDSS DR1 provides for each image
the adaptive second moments
\begin{equation}
T = M_{xx} + M_{yy} \ ,
\end{equation}
\begin{equation}
e_+ = ( M_{xx} - M_{yy} ) / T \ ,
\end{equation}
and
\begin{equation}
e_\times = 2 M_{xy} / T \ .
\end{equation}
The adaptive second moments can be converted into an axis ratio
using the relation
\begin{equation}
q_{\rm am} = \left( {1 - e \over 1 + e} \right)^{1/2} \ ,
\end{equation}
where $ e \equiv ( e_+^2 + e_\times^2 )^{1/2}$. The adaptive
moments axis ratio computed in this way are not corrected for
the effects of seeing. However, the SDSS DR1 also
provides the adaptive second moments, $T_{\rm psf}$, $e_{+,{\rm psf}}$, and
$e_{\times,{\rm psf}}$, for the point spread function at the location
of the image. These moments can be used to correct for the
smearing and shearing due to seeing; such
corrections are vital in studying the small shape changes
resulting from weak lensing \citep{be02,hi03}. However,
to eliminate the need for corrections, I will only retain
galaxies which are well resolved, with $r_e > 5 \sqrt{T_{\rm psf}}$.
Of the five bands used by SDSS, the $g$, $r$, and $i$ bands
provide useful morphological information; the low detector
sensitivity at $u$ and $z$ and the high background at $z$
make them less useful for study. At $g$, $N_{\rm gal} = 12{,}826$
galaxies satisfy my criteria, at $r$, the number is $N_{\rm gal} = 12{,}751$,
and at $i$, it's $N_{\rm gal} = 12{,}764$.

Because the weight function $w(x,y)$ is scaled
to the size of the galaxy image, the adaptive moments
axis ratio $q_{\rm am}$ can be thought of as an average axis
ratio to which the outer regions of the galaxy, beyond the
effective radius, do not contribute significantly. The
shape in the outer region can be found from the shape of
the 25 magnitude per square arcsecond isophote. The SDSS DR1 provides,
for each galaxy in each band, the values of $a_{25}$ and $b_{25}$,
the long and short semimajor axis length for the isophote
at the surface brightness 25 mag/arcsec$^2$. The isophotal
axis ratio $q_{25} \equiv b_{25}/a_{25}$ then provides a
measure of the galaxy shape in its outer regions. The
mean and standard deviation of $a_{25}/r_e$ for the
galaxies in examined in this paper is 
$2.12 \pm 0.56$ in the $g$ band,
$2.56 \pm 0.72$ in the $r$ band, and
$2.82 \pm 0.84$ in the $i$ band. 
Having two different measures of the axis ratio, $q_{\rm am}$
and $q_{25}$, thus allows comparison of the central
shape ($q_{\rm am}$) and the outer shape ($q_{25}$). Having
measures in three different bands, $g$, $r$, and $i$,
gives some information about the wavelength dependence
of the disks' apparent ellipticity. The $i$ band, having
the longest wavelength of those studied, will be least
affected by dust and by patches of star formation. 

\begin{figure}
\plotone{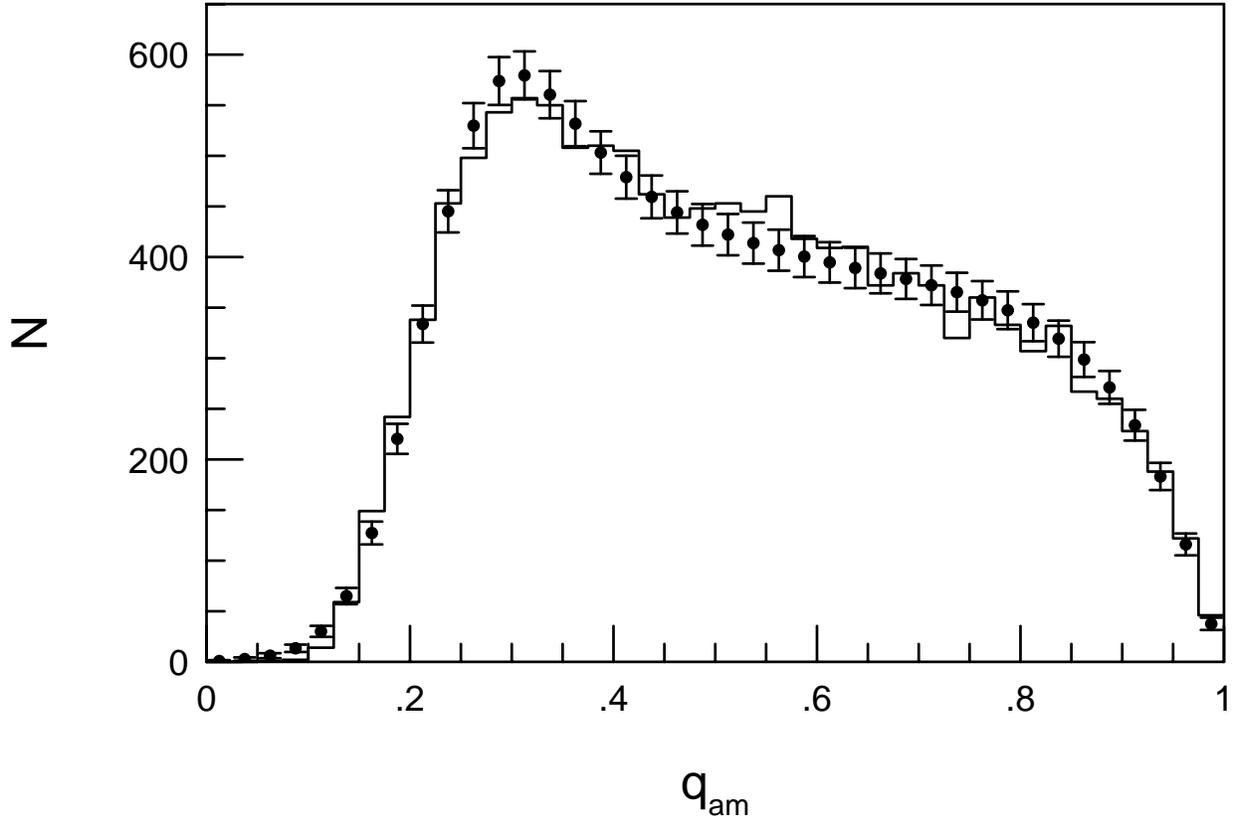}
\caption{Histogram: the distribution of axis ratio
$q_{\rm am}$, using adaptive moments in the $i$ band,
for exponential galaxies in the SDSS DR1.
Points with error bars: the best fitting model,
assuming a Gaussian distribution of disk thickness and a
log-normal distribution of intrinsic disk ellipticity. The
best fitting model has thickness $\gamma = 0.222 \pm 0.057$
and ellipticity $\ln\varepsilon = -1.85 \pm 0.89$.
}
\label{fig:q_sdss}
\end{figure}

The distribution of $q_{\rm am}$, in the $i$ band, is
shown as the histogram in Figure~\ref{fig:q_sdss}.
The relative scarcity of nearly circular
galaxies is the characteristic signature of nonaxisymmetry.
To fit the observed distribution of apparent axis
ratios, I adopted a model
in which the disk thickness $\gamma$ has a Gaussian
distribution,
\begin{equation}
f (\gamma) \propto \exp \left[ - {(\gamma-\mu_\gamma)^2 \over 2 \sigma_\gamma^2}
\right]
\label{eq:gamma}
\end{equation}
subject to the constraint $0 \leq \gamma \leq 1$, and the
disk ellipticity $\varepsilon$ has a log-normal distribution,
\begin{equation}
f (\varepsilon ) \propto {1 \over \varepsilon} \exp  \left[
- {(\ln\varepsilon - \mu)^2 \over 2 \sigma^2} \right] \ ,
\label{eq:epsilon}
\end{equation}
subject to the constraint $\ln\varepsilon \leq 0$. After assuming
values for the four parameters $\mu_\gamma$, $\sigma_\gamma$, $\mu$,
and $\sigma$, a chi-square fit to the data in Figure~\ref{fig:q_sdss}
can be made. I randomly selected a thickness
$\gamma$ and ellipticity $\varepsilon$ from the distributions in
equations~\ref{eq:gamma} and \ref{eq:epsilon}. I then randomly selected
a viewing angle $(\theta,\phi)$ and computed the resulting apparent
axis ratio \citep{bi85}
\begin{equation}
q = \left[ {A+C - \sqrt{ (A-C)^2 + B} \over
A+C + \sqrt{(A-C)^2 + B} } \right]^{1/2} \ ,
\label{eq:qform}
\end{equation}
where
\begin{eqnarray}
A = [ 1 - \varepsilon (2-\varepsilon) \sin^2 \varphi ] \cos^2 \theta
+ \gamma^2 \sin^2 \theta \ , \\
B = 4 \varepsilon^2 (2-\varepsilon)^2 \cos^2 \theta
\sin^2 \varphi \cos^2 \varphi \ , \\
C = 1 - \varepsilon (2-\varepsilon) \cos^2 \varphi \ .
\end{eqnarray}
By repeating this procedure $N_{\rm gal} = 12{,}764$ times, I created one
realization of the $\mu_\gamma$, $\sigma_\gamma$, $\mu$, $\sigma$ parameter
set. After creating 16{,}000 realizations, I computed the mean
and standard deviation of the expected number of galaxies in each of
the 40 bins in Figure~\ref{fig:q_sdss}, and computed a $\chi^2$ probability
for that particular set of parameters. A search through four-dimensional
parameter space revealed that the best fit for $q_{\rm am}$ in the
$i$ band was provided by the set of parameters
$\mu_\gamma = 0.222$, $\sigma_\gamma = 0.057$, $\mu = -1.85$, and
$\sigma = 0.89$. The resulting $\chi^2$ probability was $P = 3
\times 10^{-4}$. Formally, this is not a good fit
to the data, but more than half the contribution to $\chi^2$
comes from the bins on the far left, with $q_{\rm am} < 0.3$.
As emphasized by \citet{fa93}, the
distributions of $\gamma$ and $\varepsilon$
for spiral galaxies are almost completely decoupled,
in that the distribution of $\gamma$ determines the left hand side of $f(q)$
while the distribution of $\varepsilon$ determines the right hand side.
The model's relatively poor fit at $q_{\rm am} < 0.3$ indicates
that the galaxies in the sample have a distribution of thicknesses
which is not well fit by a Gaussian. In any case, the distribution
of disk thicknesses in my sample is not the same as the true
distribution of disk thicknesses. As shown by \citet{hu92},
a magnitude-limited sample of galaxies such as the SDSS DR1
spectroscopic galaxy sample, which has a limiting (Petrosian) magnitude
$r < 17.77$, will show a deficit of high-inclination, low-$q$
spiral galaxies. This deficit is due to extinction by dust.
In the $B$ band, for instance, an Sc galaxy is 1 to 1.5 magnitudes
fainter when seen edge-on than when seen face-on \citep{hu92}.
Although the inclination-dependent dimming is smaller at
longer wavelengths and for earlier type spirals, the fitted
disk thickness should be regarded with skepticism. Fortunately
for the purposes of this paper, spiral galaxies that appear
nearly circular ($q \gtrsim 0.8$) are nearly face-on, and
hence their thickness is almost totally irrelevant; it's
the distribution of disk ellipticity $\varepsilon$ that
determines the shape of $f(q)$ at large $q$.

\begin{figure}
\plotone{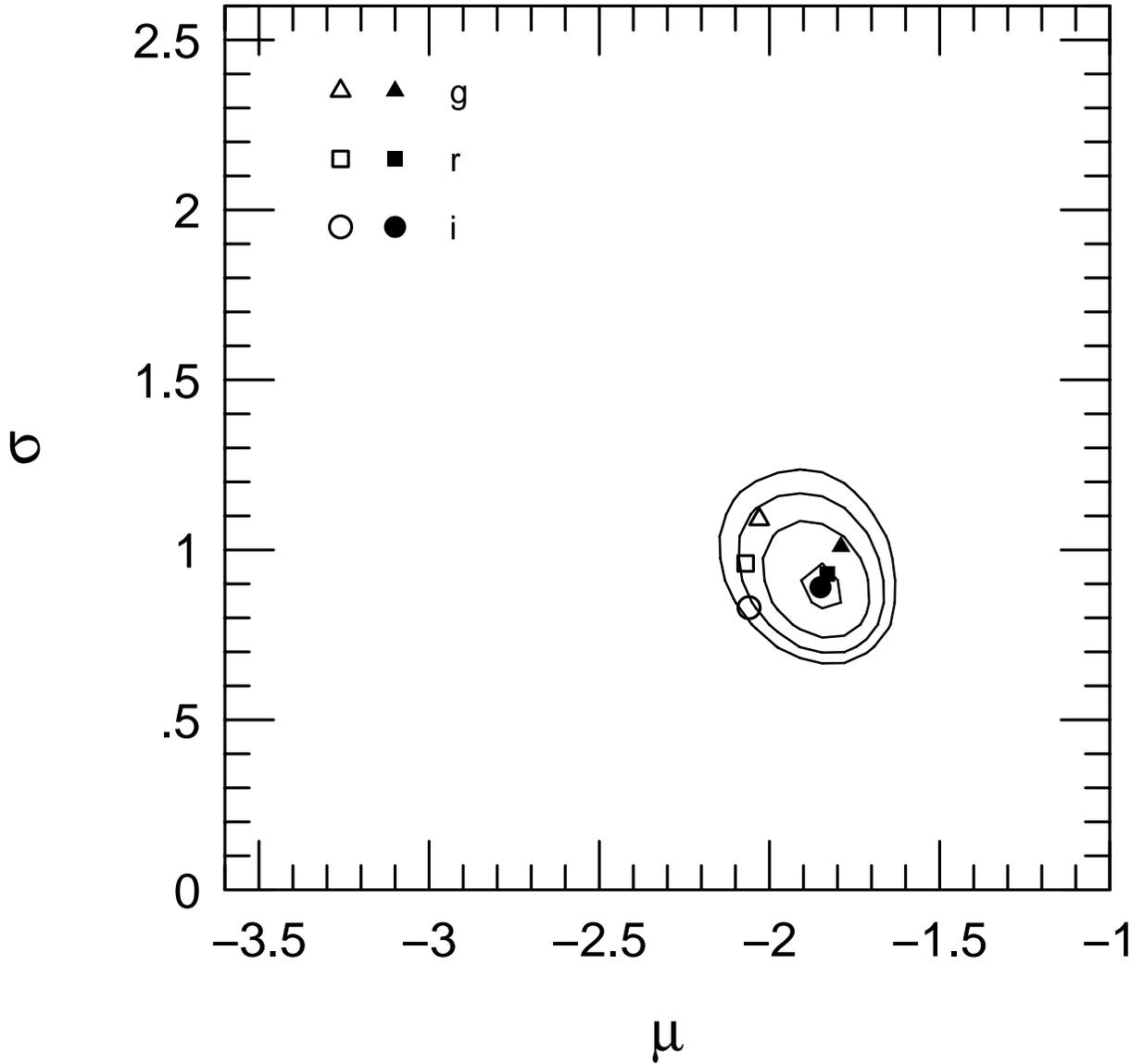}
\caption{The points indicate the best fitting values
of $\mu$ and $\sigma$, assuming a log-normal distribution
of intrinsic disk ellipticities. Open symbols use $q_{25}$,
the axis ratio of the 25 mag/arcsec$^2$ isophote, for the
apparent shape measure; filled symbols use $q_{\rm am}$, the
adaptive moments axis ratio, for the apparent shape measure.
The solid lines are isoprobability contours, as measured
by a $\chi^2$ test applied to the binned data of Figure 1.
Contours are drawn at the levels $\log_{10} P = -4$, $-6$,
$-8$, and $-10$, going from the innermost to outermost contour.
}
\label{fig:prob_sdss}
\end{figure}

Table~\ref{tab:models} shows the best fitting model
parameters, $\mu_\gamma$, $\sigma_\gamma$, $\mu$, and $\gamma$,
for the two different shape measures, $q_{\rm am}$ and $q_{25}$,
and for the three different bands, $g$, $r$, and $i$.
In addition, the best values of $\mu$ and $\gamma$ are
plotted as points in Figure~\ref{fig:prob_sdss}. Note
that going from $g$ band (triangles) to $r$ band (squares)
to $i$ band (circles) results in a smaller spread of
disk ellipticities (that is, smaller values of $\sigma$).
Going from $q_{\rm am}$ (filled symbols) to $q_{25}$ (open
symbols) results in a smaller disk ellipticity (that is,
smaller values of $\mu$). In the $i$ band, for instance,
using $q_{\rm am}$ for the shape measure results in a best fit
log-normal distribution with $\mu = -1.85$ and $\sigma = 0.89$;
the modal ellipticity for this distribution is 0.071,
the median is $\approx e^\mu \approx 0.16$, and
the mean is 0.21.
Using $q_{25}$ for the shape measure results in $\mu = -2.06$
and $\sigma = 0.83$; the modal ellipticity is 0.064,
the median is $\approx e^\mu \approx 0.13$, and
the mean is 0.17. Although the disks are rounder, on
average, in the outer regions, they still have a significant
ellipticity. Note also the color dependence of the mean
thickness $\mu_\gamma$; galactic disks are thicker, on
average, at longer wavelengths. This dependence reflects
the fact that older stellar populations, which have redder
colors, have a greater vertical velocity dispersion and
hence a greater disk scale height \citep{wi77,de98}.

The large SDSS DR1 sample of galaxies places strong constraints
on the best fitting values of $\mu$ and $\sigma$. To demonstrate
how the goodness of fit varies in $(\mu,\sigma)$ parameter
space, Figure~\ref{fig:prob_sdss} shows the isoprobability
contours for the $q_{\rm am}$ data in the $i$ band. Note
that the contours are drawn at an interval of $\Delta \log_{10}
P = 2$; the $\chi^2$ probability drops rapidly as you
move away from the best fit, indicated by the filled circle
at the center of the innermost contour. At other wavelengths,
and using the other shape measure $q_{25}$, the probability
falls off with comparable steepness in $(\mu,\sigma)$ parameter
space.

The log-normal distribution of
equation~(\ref{eq:epsilon}) is not the only functional form
to yield an adequate fit to the data. A Gaussian peaking at
$\varepsilon = 0$, as shown in equation~(\ref{eq:apm}),
provides a comparably good fit to the SDSS DR1 data. 
In the $i$ band, the best-fitting Gaussian to the $q_{\rm am}$
data has $\sigma_\varepsilon = 0.26$. The best-fitting
Gaussian to the $q_{25}$ data has $\sigma_\varepsilon = 0.21$.
The photometry provided by the SDSS DR1 is insufficient
to distinguish between a Gaussian distribution of ellipticities,
peaking at $\varepsilon = 0$, and a log-normal distribution,
peaking at $\varepsilon > 0$. Purely photometric studies, it
seems, are ill-suited to addressing the question of whether
exactly circular disks, with $\varepsilon = 0$, exist.

\section{The Andersen-Bershady Sample}
\label{andersen}

\begin{figure}
\plotone{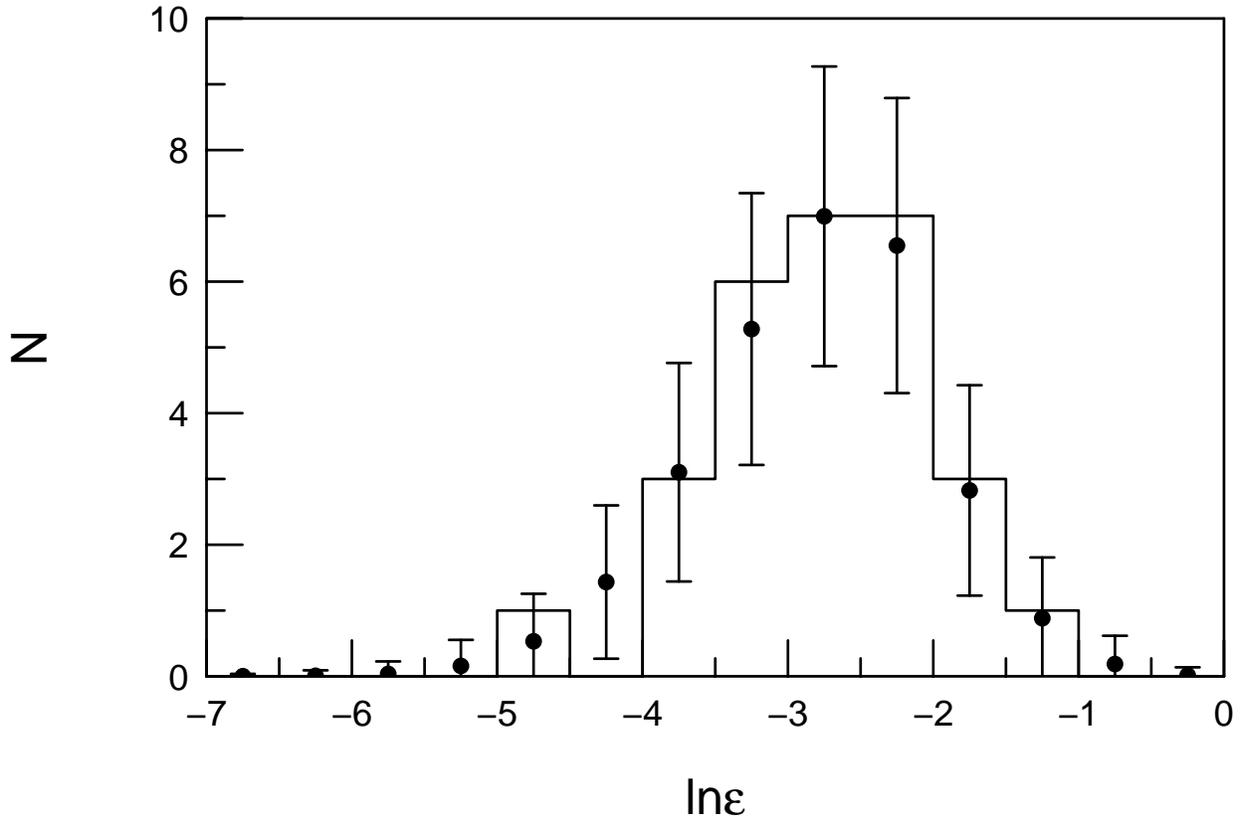}
\caption{Histogram: the distribution of
$\ln\varepsilon$ for the Andersen-Bershady
sample of galaxies. Points with error bars: the
best fitting model, assuming a log-normal distribution
of intrinsic disk ellipticity. The best fitting parent
distribution has $\ln\varepsilon = -2.29 \pm 1.04$.
(If the selection criterion $q \geq 0.866$ were
ignored, the best fit would have $\ln\varepsilon =
-2.80 \pm 0.81$.)
}
\label{fig:loge_ab}
\end{figure}

\citet{an03}, using the method outlined in \citet{an01}, combined
kinematic and $I$-band photometric data to find the disk ellipticity
for a sample of 28 nearly face-on disk galaxies; the mean
inclination of the galaxies is $i = 26^\circ$. The distribution
of $\ln\varepsilon$ determined by \citet{an03} is shown as the
histogram in Figure~\ref{fig:loge_ab}. This distribution is
well fitted by a log-normal distribution; the best fit, as
found by a $\chi^2$ test, has parameters $\mu = -2.80$ and
$\sigma = 0.81$. However, the best fit to the data in
Figure~\ref{fig:loge_ab} is not the best fit to the underlying
distribution of disk ellipticities, thanks to the selection
criteria used in building the sample. To ensure that only galaxies
with small inclination were included, \citet{an03}
selected galaxies with $q \geq 0.866$,
corresponding to an inclination $i \leq 30^\circ$ for
perfectly circular, infinitesimally thin disks.

To fit the distribution of intrinsic ellipticities
in the Andersen-Bershady sample, subject to their
selection criterion $q \geq 0.866$, I started by assuming
that the disk ellipticity has the log-normal form given
in equation~(\ref{eq:epsilon}). I further assumed, for
simplicity, that the disks are all infinitesimally thin;
since the disks in the Andersen-Bershady sample are
close to face-on, their exact thickness doesn't affect
the observed axis ratio. After assuming values for
the parameters $\mu$ and $\sigma$, I randomly
selected a disk ellipticity $\varepsilon$ as well
as a viewing angle $(\theta,\phi)$. If the resulting
apparent axis ratio, as given by equation~(\ref{eq:qform}),
was $q \leq 0.866$, I retained it in my sample. If it was
flatter than this limit, I discarded it.
By repeating this procedure until I had $N_{\rm gal} = 28$
axis ratios, I created one possible realization. After creating
16{,}000 realizations, I computed the mean and
standard deviation in each of the bins in Figure~\ref{fig:loge_ab}.
A search through parameter space revealed that the best fit,
as measured by a $\chi^2$ test, was given by $\mu = -2.29$,
$\sigma = 1.04$. The probability of the fit, illustrated by
the points and error bars in Figure~\ref{fig:loge_ab}, was
$P = 0.98$.

\begin{figure}
\plotone{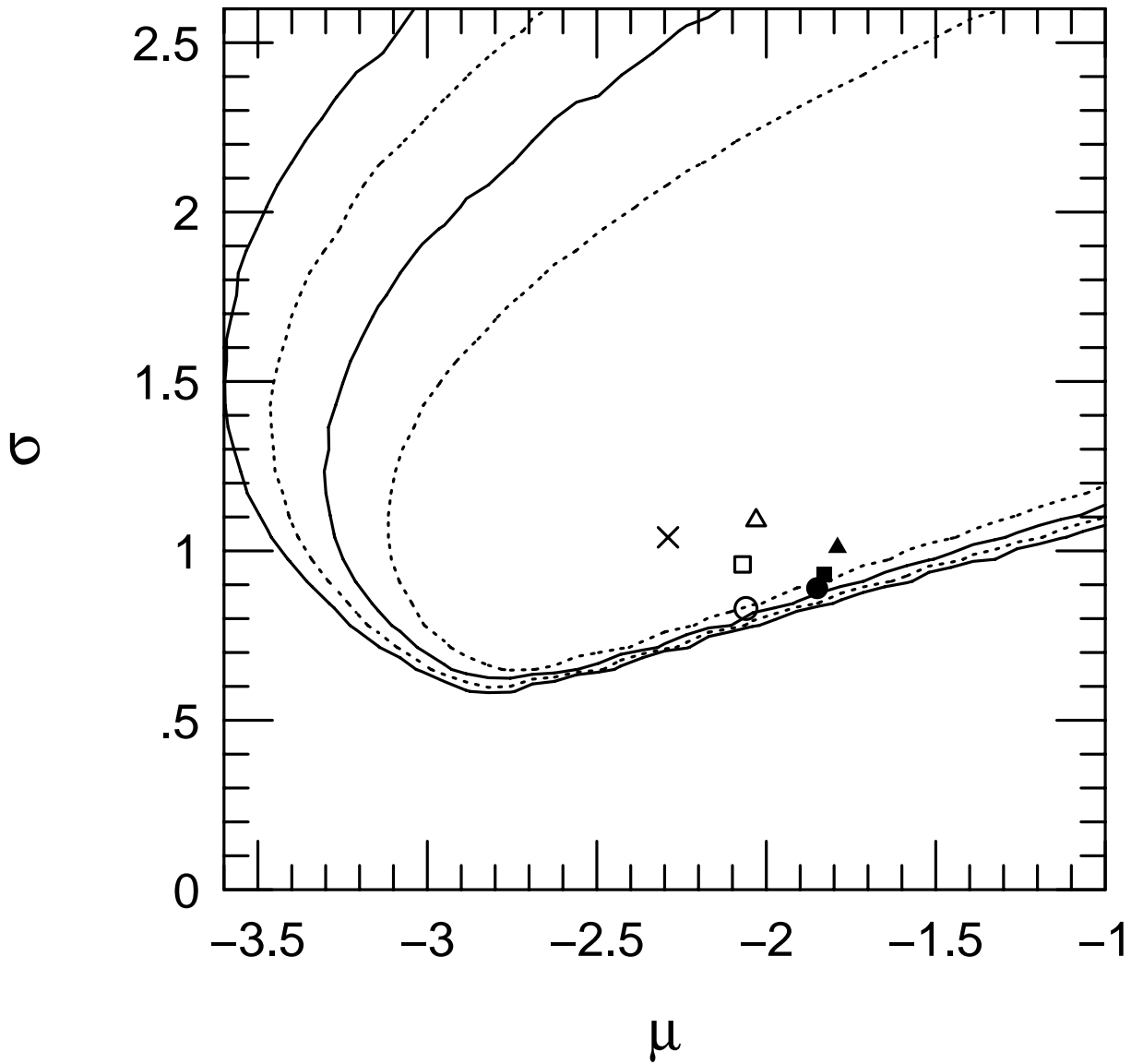}
\caption{Isoprobability contours, as measured by a $\chi^2$
test applied to the binned data of Figure 3, in ($\mu$,$\sigma$)
parameter space. Contours are drawn at the levels
$\log_{10} P = -1$, $-2$, $-3$, and $-4$, going from the
innermost to outermost contour. The cross
indicates the best fit: $\mu = -2.29$, $\sigma = 1.04$.
The best fitting points for the SDSS DR1 data are
repeated from Figure 2, for comparison purposes.
}
\label{fig:prob_ab}
\end{figure}

The 28 galaxies of \citet{an03} do not, by themselves, provide
a strong constraint on the distribution of
intrinsic ellipticities. Figure~\ref{fig:prob_ab} shows the
goodness of fit, as measured by the $\chi^2$ probability
$P_{\rm AB} (\mu,\sigma)$. The cross indicates the best fit,
and the dotted and solid lines show the isoprobability contours.
Note that the $P = 0.1$ contour --
the innermost dotted line -- encloses a large area stretching
off to the upper right of the plot. That is, a distribution
with a large value of $\mu$, corresponding to a very flattened
average shape, is acceptable as long as it is paired with
a large value of $\sigma$, signifying a wide spread in shapes.
Because the Andersen-Bershady contains only galaxies which are nearly
circular in projection, it is strongly weighted toward galaxies which
are nearly circular in their intrinsic shape, and thus cannot effectively
constrain the high-ellipticity end of $f(\varepsilon)$.

A Gaussian peaking at $\varepsilon = 0$ (see equation~(\ref{eq:apm}))
doesn't provide a good fit to the Andersen-Bershady sample. The
best-fitting Gaussian, with $\sigma_\varepsilon = 0.143$, had
a $\chi^2$ probability of only $P = 0.009$. Thus, although
the data of \citet{an03} doesn't constrain the high-ellipticity
end of $f(\varepsilon)$, its discriminatory power at low values
of $\varepsilon$ weighs strongly against a distribution peaking
at $\varepsilon = 0$.

The kinematic and photometric information exploited by
\citet{an03} is in some ways complementary to the purely
photometric information included in the SDSS DR1 axis ratios.
The nearly face-on galaxies of the Andersen-Bershady
sample constrain the low-ellipticity end of $f(\varepsilon)$;
the scarcity of nearly circular galaxies in the SDSS DR1
exponential sample (see Figure~\ref{fig:q_sdss}) constrains
the high-ellipticity end of $f(\varepsilon)$. The kinematic
measurements of \citet{an01} typically go out to $2 \to 3$
scale lengths ($1.2 \to 1.8 r_e$). The ellipticities determined
by \citet{an01} and \citet{an03} can be thought of as
average ellipticities over the inner region of the galaxy.
Thus, the Andersen-Bershady ellipticities are more directly
comparable to the ellipticities found from $q_{\rm am}$ than
from the outer axis ratios $q_{25}$. As shown in
Figure~\ref{fig:prob_ab}, the best fit using
$q_{\rm am}$ in the $i$ band, indicated by the filled
circle, is marginally consistent with the Andersen-Bershady
results. Multiplying together the probability fields
in Figure~\ref{fig:prob_sdss} and Figure~\ref{fig:prob_ab}
yields a best joint fit of $\mu = -1.89$, $\sigma = 0.96$.
For this set of parameters, the $\chi^2$ fit to the
Andersen-Bershady data has $P = 0.44$, and the fit to
the $q_{am}$ data in the $i$ band has $P = 9.5 \times 10^{-5}$.

\section{Discussion}
\label{discuss}

In this paper, I've considered two quite different
ways of determining disk ellipticities.
The SDSS DR1 axis ratios are a measure of where
the starlight is in a galaxy, and hence contain information
about bulges, bars, spiral arms, and other nonaxisymmetric
structure in the galaxies. By its nature, the analysis
of photometric axis ratios finds difficulty in distinguishing
between a Gaussian distribution of ellipticities peaking at
$\varepsilon = 0$ and a log-normal distribution peaking at
$\varepsilon > 0$. If you want to determine whether truly
circular disks actually exist, examining the apparent axis
ratios of disks (even of a large number of disks) is not
an effective method to use. By contrast, the approach of
\citet{an01} and \citet{an03} uses both photometric and
kinematic information to find the ellipticity of individual
disks. Since \citet{an03} looked at disks which are nearly
circular in projection, their data are ineffective at determining
the high-ellipticity end of $f(\varepsilon)$.

Although both data sets are reasonably well fitted by a log-normal
distribution with $\ln\varepsilon = -1.89 \pm 0.96$, this distribution
should not be engraved on stone as {\em the} distribution of
disk ellipticities in spiral galaxies. Describing a complex
structure such as a spiral galaxy with a single number $\varepsilon$
(or even two numbers, $\varepsilon$ and $\gamma$) requires
averaging over a great deal of substructure. How one takes the
average will affect the value of $\varepsilon$ found. For
instance, using the adaptive moments shape $q_{\rm am}$ yields
larger values for the ellipticity than using the isophotal
shape $q_{25}$. Moreover, observations at different wavelengths
result in different values of $\varepsilon$.

If a disk of gas and stars is orbiting in a logarithmic
potential which is mildly elliptical in the disk plane,
with $\varepsilon_\phi \ll 1$, then the integrated line
profile from the disk will have a width $W = 2 v_c
(1 - \varepsilon_\phi \cos 2 \varphi ) \sin \theta$ when
viewed from a position angle $\varphi,\theta$ \citep{fr92}.
The alteration in the line width, due to noncircular motions
in the elliptical potential, will produce a scatter in
the observed Tully-Fisher relation between line width and
absolute magnitude \citep{tu77}. For an ellipticity
$\varepsilon_\phi = 0.1$, the expected scatter is 0.3
magnitudes \citep{fr92}. (This assumes that the inclination
has been determined accurately using kinematic information;
if the inclination is determined photometrically, assuming
the disk is circular, there will be an additional source
of scatter.)

\begin{figure}
\plotone{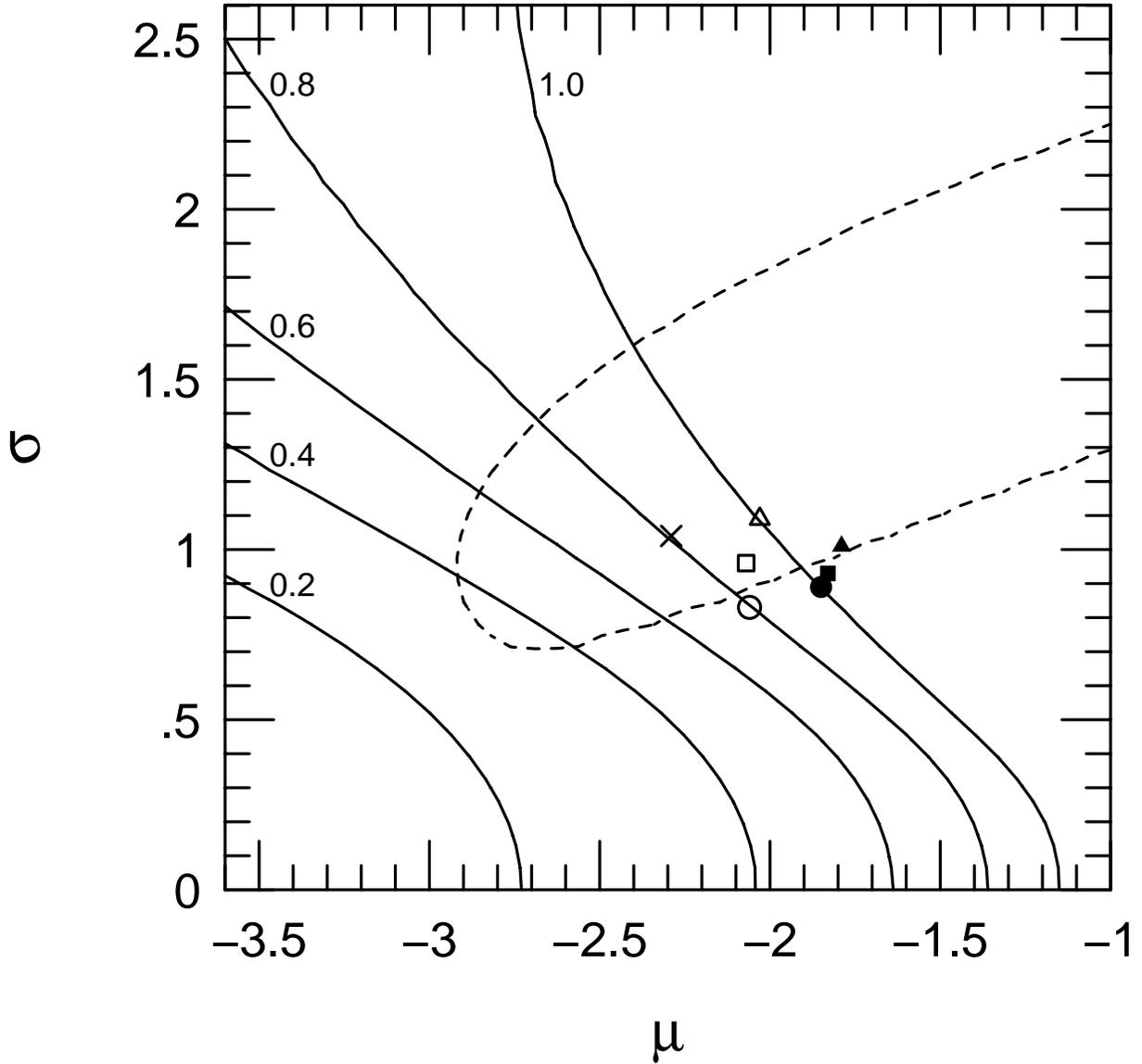}
\caption{The scatter, in magnitudes, around the
mean Tully-Fisher relation, for a log-normal distribution
of potential ellipticity. The contours are drawn, starting
at the lower left, at the levels 0.2, 0.4, 0.6, 0.8, and 1.0
magnitudes. The dashed line indicates the 50\% probability
contour for the Andersen-Bershady data. The cross is the
best fit for the Andersen-Bershady data. The open and filled
symbols are the best fits for the SDSS DR1 data, repeated
from Figure 2.}
\label{fig:scatter_tf}
\end{figure}

If the potential ellipticity $\varepsilon_\phi$
is assumed to be drawn from a log-normal distribution,
and the embedded disk is viewed from a random angle, the
resulting scatter in the Tully-Fisher relation is shown
in Figure~\ref{fig:scatter_tf}. The best fits for
the SDSS DR1 data are superimposed as triangles ($g$ band),
squares ($r$ band) and circles ($i$ band). Even in the
$i$ band, the best fit to the ellipticity of the disks
would produce far more scatter than is seen in the
Tully-Fisher relation. Using $q_{\rm am}$ as the shape
measure (as shown by the filled circle in
Figure~\ref{fig:scatter_tf}),
1.0 magnitude of scatter is predicted. Using
$q_{25}$ as the measure (as shown by
the open circle in Figure~\ref{fig:scatter_tf}),
0.8 magnitudes of scatter is predicted.
The best fit to the Andersen-Bershady data,
shown as the cross in Figure~\ref{fig:scatter_tf}, would
also produce 0.8 magnitudes of scatter in the Tully-Fisher
relation.

In contrast, the actual scatter in the Tully-Fisher
relation is smaller than 0.8 magnitudes. \citet{co97}
found 0.46 magnitudes of scatter in the optical
Tully-Fisher relation when he used, as his velocity
measure, the rotation speed at 2.2 scale lengths
($\sim 1.3 r_e$), about the extent of the kinematic
data of \citet{an01}. \citet{ve01}, in his study
of spiral galaxies in the Ursa Major cluster, found
still smaller scatters. Combining near-infrared $K'$
magnitudes with $V_{\rm flat}$, the rotation speed
in the outer, flat part of the rotation curve, \citet{ve01}
found a best fit with zero intrinsic scatter, with an
upper limit, at the 95\% confidence level, of 0.21 magnitudes.

The SDSS DR1 data are clearly inconsistent with such
small scatters in the Tully-Fisher relation. The
light distribution in the $i$ band cannot reflect
the ellipticity of the underlying potential, but
must be due primarily to nonaxisymmetric structures
such as bars, spiral arms, non-circular rings, and
so forth. It should be noted that lopsidedness
($m = 1$ distortions) is not uncommon in disk galaxies;
\citet{ri95} found that a third of the galaxies in
their sample of nearly face-on spirals had significant
lopsidedness at 2.5 scale lengths ($\sim 1.5 r_e$).
Unfortunately, the SDSS DR1 does not provide, among
its tabulated parameters, the odd moments that would
permit a quantitative estimate of disk lopsidedness.

Although barred galaxies were excluded
from the Andersen-Bershady sample, I made no
effort to sift out barred galaxies from my
SDSS DR1 sample. The Andersen-Bershady
results are not inconsistent with a small scatter
in the Tully-Fisher relation. The dashed line in
Figure~\ref{fig:scatter_tf} is the $P = 0.5$ contour
for the Andersen-Bershady sample; that is, every
$(\mu,\sigma)$ pair within the dashed line gives
a ``too-good-to-be-true'' fit to the sample of
\citet{an03}. This contour encloses $(\mu,\sigma)$
pairs which produce as little as 0.32 magnitudes
of scatter. At lower probability levels, the $P = 0.1$
contour yields as little as 0.28 magnitudes and
the $P = 0.01$ contour yields as little as 0.25
magnitudes of scatter. In summary, the Andersen-Bershady
data are consistent with as little as a quarter-magnitude
of scatter in the Tully-Fisher relation. The
Andersen-Bershady data are also consistent with
the adaptive moments axis ratios from the SDSS DR1.
However, the three sets of information -- the
Andersen-Bershady data, the Tully-Fisher scatter
(or lack thereof), and the SDSS DR1 axis ratios --
are not mutually consistent. The SDSS DR1 axis
ratios, if they accurately traced the potential
ellipticity, would produce too much scatter in
the Tully-Fisher relation.

\acknowledgments

I thank Matt Bershady, Dave Andersen,
Richard Pogge, and Albert Bosma
for their helpful and courteous comments.
Ani Thakar was an invaluable guide to the sdssQA
query tool.
The Sloan Digital Sky Survey (SDSS) is a joint project of
The University of Chicago, the Institute of Advanced Study,
the Japan Participation Group, the Max-Planck-Institute for
Astronomy (MPIA), the Max-Planck-Institute for Astrophysics (MPA),
New Mexico State University, Princeton Observatory, the
United States Naval Observatory, and the University of Washington.
Apache Point Observatory, site of the SDSS telescopes, is operated
by the Astrophysical Research Consortium (ARC). Funding for the
project has been provided by the Alfred P. Sloan Foundation,
the SDSS member institutions, the National Aeronautics and
Space Administration, the National Science Foundation, the
U.S. Department of Energy, the Japanese Monbukagakusho, and
the Max Planck Society. The SDSS website is http://www.sdss.org/.

\clearpage
\begin{deluxetable} {c c c c c c}
\tablewidth{0pt}
\tablecaption{Best Fitting Models: Gaussian Thickness
Distribution, Log-normal Ellipticity Distribution\label{tab:models}}
\tablehead{
\colhead{Shape Measure} & \colhead{Band} &
\colhead{$N_{\rm gal}$} & \colhead{$\mu_\gamma \pm \sigma_\gamma$} &
\colhead{$\mu \pm \sigma$} & \colhead{$P_{\chi^2}$} }
\startdata
                 & $g$ & 12,826 
 & $0.205 \pm 0.054$ & $-1.79 \pm 1.01$ & $5 \times 10^{-5}$ \\
Adaptive moments & $r$ & 12,751
 & $0.216 \pm 0.056$ & $-1.83 \pm 0.93$ & $5 \times 10^{-4}$ \\
 ($q_{\rm am}$)  & $i$ & 12,764 
 & $0.222 \pm 0.057$ & $-1.85 \pm 0.89$ & $3 \times 10^{-4}$ \\
\hline
                           & $g$ & 12,826
 & $0.211 \pm 0.056$ & $-2.03 \pm 1.09$ & $2 \times 10^{-2}$ \\
25 mag/arcsec$^2$ isophote & $r$ & 12,751
 & $0.231 \pm 0.064$ & $-2.07 \pm 0.96$ & $3 \times 10^{-3}$ \\
      ($q_{25}$)           & $i$ & 12,764
 & $0.248 \pm 0.074$ & $-2.06 \pm 0.83$ & $2 \times 10^{-6}$ \\
\hline
Andersen-Bershady & $I$ & $28$ & & $-2.29 \pm 1.04$ & $0.996$ \\
\enddata
\end{deluxetable}

\clearpage

\clearpage

\clearpage

\clearpage

\clearpage

\end{document}